# Review on Decarbonizing the Transportation Sector in China: Overview, Analysis, and Perspectives

Jiewei Li[1,2], Ling Jin[3], Han Deng[2], Lin Yang[1*]
1 Urban Governance and Design Thrust, The Hong Kong University of Science and Technology (Guangzhou), Guangzhou, China (*Corresponding Author)
2 Institute for Transportation and Development Policy, Guangzhou, China
3 Innovation, Policy and Entrepreneurship Thrust, The Hong Kong University of Science and Technology (Guangzhou), Guangzhou, China

**Abstract**
This review identifies challenges and effective strategies to decarbonize China's rapidly growing transportation sector, currently the third largest carbon emitter, considering China's commitment to peak carbon emissions by 2030 and achieve carbon neutrality by 2060. Key challenges include rising travel demand, unreached peak car ownership, declining bus ridership, gaps between energy technology research and practical application, and limited institutional capacity for decarbonization. This review categorizes current decarbonization measures, strategies, and policies in China's transportation sector using the "Avoid, Shift, Improve" framework, complemented by a novel strategic vector of "Institutional Capacity & Technology Development" to capture broader development perspectives. This comprehensive analysis aims to facilitate informed decision-making and promote collaborative strategies for China's transition to a sustainable transportation future.

## 1   Introduction

China's commitment to the "dual carbon" goal - peaking carbon emissions by 2030 and achieving carbon neutrality by 2060 - requires transformative changes across all sectors of the economy. The transportation sector, which contributes 9% of the country's total carbon emissions and is the fastest growing source of emissions[1], is a critical area for action. Despite progress in developing high-speed rail and promoting electric vehicles during the 13th Five-Year Plan, much of the transportation system still relies on carbon-intensive methods[2]. This underscores the need for China's 14th Five-Year Plan goals such as increasing the share of public transportation in cities, reducing vehicle energy consumption, and promoting green logistics systems[3]. In this article, we present an analytical framework for China's transport decarbonization strategy, focusing on the "Avoid, Shift, Improve" (ASI)[4] strategy vector, which includes measures to decarbonize both passenger and freight transport. In addition, we use the novel "Institutional Capacity & Technology Development" (ICTD) strategy vector to highlight the importance of governance capacity, cross-sector collaboration, and emerging technologies. We explore how these strategies could be implemented in China's unique context, discussing specific measures, analyzing their potential impacts, and providing successful examples from China and abroad, with the aim of contributing to the ongoing dialogue on effective policy tools and strategies to achieve China's ambitious carbon neutrality goal.

## 2   Methods

This review examines the decarbonization of China's transportation sector using a thematic analysis. First, we established a boundary using a Tank-to-Wheel approach[5], focusing on emissions from fuel combustion during vehicle operation, while also highlighting key strategies in manufacturing, fuel and energy production to align with China's policy focus. Data was collected from various policy documents, reports, and articles from government and academic databases. These were categorized under the ASI framework and the novel new vector: "Institutional Capacity & Technology Development" to assess policy implementation capacity and technological advancement. Each strategy within these frameworks was analyzed for effectiveness, challenges, and areas for improvement. Our findings were then contextualized to provide suggestions for improving and adopting strategies to promote decarbonization.

## 3  Results and Discussion

### 3.1  Strategy Vector I: Avoid, Shift, Improve

The ASI strategy serves as a comprehensive analytical framework for reducing carbon emissions in the transportation sector. This three-pronged approach includes avoiding unnecessary travel, shifting to less carbon-intensive modes, and improving energy efficiency.

#### 3.1.1 Avoid

The Avoid strategy, aimed at reducing travel demand, curbing carbon emissions at their source, necessitating tailored implementation considering China's unique urban and societal context. For example, mixed-use development, adopted in rapidly urbanizing areas including high-tech parks and industrial areas around Shanghai and Beijing[6], reduces commuting distances by promoting residence-occupation adjacency. In comparison, spatial clustering allows development of efficient transportation strictures and therefore limits commuting[7]. Regarding digital technologies, online collaboration platforms like DingTalk reduce the need for commuter travel. Yet, to enhance this trend, policymakers should incentivize remote work adoption through tax benefits and supportive regulations for home-based work while ensuring robust digital infrastructure. Travel demand management measures, like traffic and license plate restrictions, can disincentivize high-carbon activities and avoid unnecessary car trips. Thus, while the Avoid strategy implementation presents challenges, it also offers distinctive opportunities to pioneer innovative, context-specific measures for the decarbonization of transport in China.

#### 3.1.2 Shift

The "Shift" strategy involves shifting from high-carbon modes of transportation like private cars and fossil-fuel-dependent freight trucks, to more sustainable, low-carbon alternatives, including public transit, non-motorized transport (NMT), freight trains, and shared mobility solutions. These modes offer significant opportunities to reduce carbon emissions, given the right infrastructure, regulatory support and public acceptance. China is at the forefront of this shift with its extensive public transportation network and world-leading bike-sharing program. However, realizing the full potential of this strategy will require an integrated approach that combines planning, regulatory support, and public participation while addressing the diverse needs of different user groups.

**(1) Public Transit Improvement:** Public transit systems in many Chinese cities have witnessed notable improvements and expansions. The operating lengths of China's urban rail transit has expanded from 1,699 km in 2011 to 8,735.6km in 2021[8], providing affordable, efficient and low-carbon alternatives to private cars. Despite these efforts, public transit use can be further optimized through measures such as improving connectivity between different transit modes, designate bus lanes, operate collector routes or branch lines to expand the service area, etc. Mobility-as-a-Service (MaaS) platforms can also be further utilized to provide on-demand bus service and help make public transit more accessible and user-friendly, fostering a shift away from private vehicle use.

**(2) Non-Motorized Transportation (NMT) Improvements:** NMT improvements promote walking and cycling as alternative modes of travel, incentivized by ensuring safety and improving user experience. Cities like Beijing and Yichang have developed NMT development plans, stressing its contribution to decarbonization. Synergy between NMT and public transport is essential[9], with continuous pedestrian walkways, protected bike lanes, and dedicated bike parking near transit stations. Updating e-bike policies and regulations could facilitate their use in urban areas while ensuring road safety. Streets and intersections, especially in areas of high commercial activity, should be redesigned to calm vehicle traffic and provide more space for pedestrians and bicycles. For long-term sustainability and effectiveness, integrating NMT into urban planning and design from the outset through compact urban development and the design of urban spaces around NMT and public transport is critical.

**(3) Travel Demand Management (TDM):** TDM aims to disincentivize private car use and thus encourage shifting to low-carbon modes and avoid unnecessary car trips, through measures like car restrictions, low-emission zones, congestion pricing, parking management, and high occupancy vehicle (HOV) lane promotion. Shanghai, Beijing and Guangzhou have implemented license plate lotteries and auctions to control new vehicle numbers to control new vehicle numbers[10]. License plate number-based restrictions have applied in many cities to reduce traffic during peak hours. The creation of low-emission zones in city centers, restricting high-polluting vehicles, is an innovative strategy being explored, which has seen successful implementation in Jinan[11]. Congestion pricing has seen success in places like Singapore[12] and could alleviate Chinese cities' peak-hour traffic. Parking management strategies like price differentiation and reducing on-street parking can also disincentivize private car use while making room for non-motorized traffic. Effective enforcement and compliance, institutional capacity, public awareness, and regulatory mechanisms are crucial for success.

**(4) Integrating Urban Public Transit with Intercity Transport:** Seamless integration of urban public transit and intercity systems, including cross-platform transfer, short transfer distance and unified fares [13], can significantly reduce the friction in transferring between public transit and intercity transport, lessen dependence on private vehicles and lowering carbon emissions. Coordination of schedules and streamlined transit hubs further aid this process.

**(5) Promoting Multimodal Freight Transport:** Transitioning from road-centric to multimodal freight strategies is key to decarbonizing freight transport, given the higher unit carbon emissions of road freight compared to rail and waterways[14]. As seen in Hubei Province, integrating road, rail, and waterway logistics means harmonizing modes of transport to leverage their environmental efficiencies. Strategies such as road-to-rail and road-to-water transfers, as well as single order throughput, help reduce freight's carbon footprint. This optimization requires efficient transshipment facilities, harmonized regulations, and advanced freight handling technology.

*3.1.3 Improve*

The Improve strategy focuses on enhancing energy efficiency and reducing the carbon intensity of transportation modes, notably through electrification. It encompasses not only technological innovations but also the optimization of current systems to boost their efficiency.

**(1) Energy Efficiency Improvements & New Technology:** This measure emphasizes the adoption of improved fuel economy standards, low-carbon operational vehicles, autonomous vehicles, and hydrogen fuel cell technology. For instance, China has progressively tightened its fuel economy standards for passenger cars[15], contributing to the reduction of average fuel consumption. Further enforcement and advancement of these standards, along with the promotion of new technologies, could see significant reductions in transportation-related emissions.

**(2) Electrification:** Electric vehicles (EVs), including private cars, taxis, buses, public service vehicles, and freight trucks, have experienced a significant upswing in China. For private cars, China's extensive investment in charging infrastructure, consumer subsidies, and the green license plate system have spurred this growth. By 2023, China's new energy vehicle sales is expected reach 8.5 million units, accounting for 36% of total vehicle sales[16]. For public buses, new energy bus has accounted for 59% fleet and 96% of sales in 2019, where 90.1% of them are pure electric buses[17]. As purchase subsidies are gradually phased out, China is focusing on expanding charging infrastructure by setting allocation standards and providing operating incentives to promote the use of electric vehicles. Importantly, as China's energy generation remains primarily thermal-based with high carbon emission factors, reducing fossil fuel use in power generation is crucial to the effectiveness in decarbonizing transportation through electrification[18]. Designing peak-shaving incentive mechanisms for vehicle-to-grid technology[19] and shifting charging towards low-carbon hours[20], establishing renewable energy charging stations can help reduce use of fossil fuel-generated electricity, while exploring alternatives for battery materials, reuse and recycling options for used batteries, can help reduce the lifecycle carbon emissions of EVs.

**(3) Maximizing System Efficiency:** Digital management systems have immense potential to improve the efficiency of transportation systems and reduce carbon emissions. Application of intelligent transportation system[21] has demonstrated how digitalization can help regulate traffic flow and reduce congestion, thereby reducing carbon emissions. In addition, the rise of 5G, automated driving technology, and vehicle-road coordination systems promises to minimize traffic flow disruptions and fuel consumption associated with constant stop-start movements[22]. By optimizing traffic dynamics, such cutting-edge systems make a significant contribution to reducing carbon emissions from transportation.

### 3.2 Strategy Vector II: Institutional Capacity & Technology Development
#### 3.2.1 Institutional Capacity

Decarbonizing the transportation sector is a critical component of China's efforts to reduce its carbon emissions and meet its climate goals. To ensure the success of these efforts, it is essential to build and strengthen institutional capacity through harmonized carbon accounting standards, reliable monitoring, reporting and verification (MRV) systems, and effective cross-sectoral cooperation mechanisms.

**(1) Harmonizing Carbon Accounting Standards:** China has made significant progress in developing carbon accounting standards for its transportation sector at the national and regional levels, guided by the National Development and Reform Commission's (NDRC) Greenhouse Gas Accounting and Reporting Guidelines[23]. Pioneering efforts, such as Beijing's 2013 pilot emissions trading scheme (ETS), have resulted in a comprehensive carbon accounting system that provides valuable lessons for expanding the national ETS. At the same time, China's national ETS trading system requires companies to establish monitoring, reporting, and verification (MRV) systems to ensure accurate reporting of carbon emissions. The China Certified Emission Reductions (CCER) mechanism facilitates the trading of emission reduction credits, as seen in the Shenzhen ETS launched in 2013. Looking ahead, a unified GHG accounting framework that integrates best practices and carbon accounting standards in transportation infrastructure planning is needed. Improved MRV systems will ensure accurate data reporting, and digital technologies such as blockchain can improve the transparency and efficiency of CCER.

**(2) Long-term Roadmap and Cross-sectoral Cooperation:** Developing a decarbonization roadmap and establishing cross-sectoral cooperation mechanisms is critical for China's transport sector[24]. The roadmap should be citizen-centered and tailored to the local status quo, supported by policy simulations and studies that assess carbon reduction impacts through a robust carbon accounting system. Shanghai and Guangzhou have formulated carbon reduction tasks in several aspects, with short- and long-term objectives. After developing a roadmap, cross-sectoral cooperation involving various government and private sectors is essential for informed decision-making and effective implementation of decarbonization measures. Specialized teams and cross-city mechanisms can facilitate information sharing, technical discussions, and project coordination, reducing administrative barriers and enabling comprehensive decarbonization of the transportation sector.

#### 3.2.2 Technology Development

**(1) Hydrogen Technology:** China's 13th Five-Year Plan recognized hydrogen as a strategic industry and promoted research, infrastructure development, and commercialization of hydrogen fuel cell vehicles (FCVs)[25]. Currently, two technologies dominate the scene: hydrogen fuel cell technology, which offers superior energy density and robust low-temperature resistance, making it ideal for year-round, long-distance transportation; and hydrogen internal combustion engine technology, which has achieved an effective thermal efficiency of 41.8%, surpassing gasoline counterparts[26]. The National Development and Reform Commission aims to have about 50,000 FCVs on the road by 2025[27]. International examples also underscore the potential of hydrogen technology. In its 2017 Basic Strategy for Hydrogen Energy, Japan aims to expand hydrogen refueling stations and increase the number of hydrogen fuel cell vehicles and buses. Meanwhile, the United States has released legislation and plans to moderate the price of hydrogen and further support hydrogen R&D. In China, the hydrogen industry still faces challenges such as high production costs, inadequate infrastructure, and limited market demand. The government should utilize policies such as R&D funding, tax incentives, and purchase incentives for new hydrogen vehicles to catalyze leapfrog advances and spur growth in China's hydrogen industry.

**(2) Aviation Decarbonization:** In 2019, CO2 emissions from China's aviation sector accounted for 1% of total carbon emissions, with 79% originating from in-flight combustion and 20% from ground operations[28]. Combustion of 1 kg of sustainable aviation fuel (SAF) can mitigate at least 2.205 kg of CO2 emissions[29]. At the end of 2022, the first commercial cargo flight using SAF has completed its journey in mainland China. The 14th Five-Year Plan for the green development of civil aviation projects that China will use 50,000 tons of SAF by 2025. Despite its potential, the current high cost of SAF makes advanced biofuels a more immediate alternative. These fuels can be produced from sources such as vegetable oils, waste oils, gasified household waste, or the fermentation of wood residues. Synthetic fuels derived from hydrogen and captured carbon could become a scalable solution, with the $CO_2$ potentially sourced from direct air capture technology powered by renewable electricity and then used alongside water and renewable electricity to produce hydrogen and synthetic gas[30].

## 4  Conclusions

This study analyzes China's decarbonization strategy for its rapidly expanding transportation sector through the lens of the ASI and ICTD strategy vectors, recognizing the need for robust governance, cross-sector collaboration, and tailored adaptations to China's unique context, offering an important blueprint for a sustainable transportation future with global resonance. While providing important insights, the analysis has limitations, focusing primarily on carbon emissions from fuel combustion without considering the full life cycle, and the context-specific nature of China may limit global applicability. The dynamic nature of technology and policy landscapes underscores the need for ongoing strategic updates, and each city and country should adapt and weigh these strategies differently in accordance with their unique transportation and industrial characteristics.

## Acknowledgements

We extend gratitude to Prof. Lin Yang at HKUST(GZ) for her supervision, Xianyuan Zhu at ITDP and Yonghao Xu at HKUST(GZ) for their expertise, and to Red Bird MPhil, HKUST(GZ) for enabling this work financially.